\documentclass[aps,showpacs,nofootinbib,superscriptaddress]{revtex4}

\usepackage{graphicx}
\usepackage{bm}
\usepackage{amsmath}
\usepackage{amssymb}
\usepackage{hyperref}

\newcommand{\g}{\gamma}

\newcommand{\qslash}{\kern 0.2 em n\kern -0.50em /}
\newcommand{\nslash}{\kern 0.2 em n\kern -0.50em /}
\newcommand{\kslash}{\kern 0.2 em k\kern -0.45em /}
\newcommand{\lslash}{\kern 0.2 em l\kern -0.50em /}
\newcommand{\pslash}{\kern 0.2 em p\kern -0.50em /}
\newcommand{\Sslash}{\kern 0.2 em S\kern -0.50em /}
\newcommand{\Pslash}{\kern 0.2 em P\kern -0.50em /}
\newcommand{\Dslash}{\kern 0.2 em D\kern -0.65em /\kern 0.15em}
\newcommand{\slim}{\mskip 1.5mu}

\newcommand{\eps}{\epsilon}

\newcommand{\Tr}{\operatorname*{Tr}\nolimits}

\newcommand{\ii}{i}

\usepackage{color}

\begin{document}

\title{Twist-3 T-odd fragmentation functions $G^\perp$ and $\tilde{G}^\perp$ in a spectator model}

\author{Yongliang Yang}\affiliation{Department of Physics, Southeast University, Nanjing
211189, China}
\author{Zhun Lu}\email{zhunlu@seu.edu.cn}\affiliation{Department of Physics, Southeast University, Nanjing
211189, China}
\author{Ivan Schmidt}\email{ivan.schmidt@usm.cl}\affiliation{Departamento de F\'\i sica, Universidad T\'ecnica Federico Santa Mar\'\i a, and
Centro Cient\'ifico-Tecnol\'ogico de Valpara\'iso,
Casilla 110-V, Valpara\'\i so, Chile}

\begin{abstract}
We present a calculation of the twist-3 T-odd chiral-even fragmentation functions $G^{\perp}$ and $\tilde{G}^{\perp}$ using a spectator model. We consider the effect gluon exchange to calculate all necessary one-loop diagrams for the quark-quark and quark-gluon-quark correlation functions. We find that the gluon loops corrections generate non-zero contribution to these two fragmentation function. We numerically calculate their half-$k_T$ moments by integrating over the transverse momentum and also verify the equation of motion relation among $G^{\perp}$, $\tilde{G}^{\perp}$ and the Collins function.
\end{abstract}

\pacs{13.60.Le,13.87.Fh,12.39.Ki}

\maketitle

\section{Introduction}

Our understanding of the hadron structure depends on what we know about the parton distribution functions and fragmentation functions.
These functions appear in the decompositions of the parton correlation functions.
In recent years, several kinds of experiments have been carried out, such as the semi-inclusive deep inelastic scattering (SIDIS), and $e^+e^-$ annihilation into hadrons, which provide us considerable information on a class of T-odd and chiral-odd fragmentation functions.
A notable example is the Collins function~\cite{Collins:1992kk} $H_1^\perp$ that describes the fragmentation of a transversely polarized quark to an unpolarized hadron (e.g. a pion) and can be used to analyse the hadronic quark spin contribution.
It has been widely recognized that the Collins function plays an important role in the understanding of the transverse single spin asymmetries (SSAs).
In addition, the $e^+\,e^-$ annihilation data combined with the SIDIS data can be applied to extract~\cite{Anselmino:2013vqa,Kang:2015msa} the Collins function.

Within the field theoretical framework of QCD, there are two approaches to interpret SSAs in high energy processes: the transverse-momentum-dependent (TMD) approach~\cite{Collins:1992kk,Sivers:1989cc,Boer:2003cm} and the twist-3 collinear factorization in terms of multi-parton correlation~\cite{Qiu:1991wg,Qiu:1998ia,Kang:2008ih}.
Recently, it was suggested~\cite{Metz:2012ct} that the fragmentation contribution in the twist-3 collinear framework may be also important for the SSA in pp collision.
Later phenomenological analysis~\cite{Kanazawa:2014dca} showed that, besides the contribution of the twist-3 collinear distribution functions, twist-3 fragmentation functions are also necessary for describing the SSA data in both SIDIS and $pp$ collision~\cite{Adams:2003fx,Abelev:2008af,Adamczyk:2012xd,Lee:2007zzh} in a consistent manner~\cite{Kang:2011hk}.
Three chiral-odd fragmentation functions, $\hat H(z)$, $H(z)$ and $\hat H_{FU}^{\Im}(z,z_1)$, participate in those processes.
The first one corresponds to the first $k_T$-moment of the TMD Collins function and has been applied to interpret the SSA in $pp$ collisions in previous studies~\cite{Yuan:2009dw,Kang:2010zzb}.
The second one appears in the subleading order of a $1/Q$ expansion of the quark-quark correlator, while its TMD version $H(z,k_T^2)$ is also a twist-3 function.
The function $\hat H_{FU}^{\Im}(z,z_1)$ is the imaginary part of $H_{FU}(z,z_1)$, and is connected to
another fragmentation function $\tilde{H}$(z) through an integration over $z_1$~\cite{Metz:2012ct,Lu:2015wja},
with $\tilde{H}(z)$ the collinear version of $\tilde{H}(z,k_T^2)$, which is encoded in the TMD quark-gluon-quark correlation function.
It has been found that $\tilde{H}$ also plays an important role in the transverse SSA $\sin\phi_S$ in SIDIS~\cite{Wang:2016tix}.

At the twist-3 level, apart from $\tilde{H}(z,k_T^2)$ and $H(z,k_T^2)$, there are two other T-odd TMD fragmentation functions for a spin-0 hadron, denoted by $\tilde{G}^\perp(z, k_T^2)$ and ${G}^\perp(z, k_T^2)$.
They appear in the decompositions of the quark-gluon-quark and quark-quark correlators, respectively.
In the TMD framework~\cite{Bacchetta:2006tn}, the fragmentation function $\tilde{G}^\perp(z,k_T^2)$ may give rise to longitudinal beam SSA (denoted by $A_{LU}^{\sin\phi}$) and target SSA (denoted by $A_{UL}^{\sin\phi}$, through the coupling with the unpolarized distribution function $f_1$.
Sizable longitudinal SSAs in SIDIS have already been measured by the HERMES~\cite{Airapetian:2005jc,Airapetian:2006rx}, CLAS~\cite{Avakian:2003pk,Aghasyan:2011ha,Gohn:2014zbz} and COMPASS~\cite{Alekseev:2010dm} Collaborations.
The SSAs are phenomenologically interpreted by the twist-3 distribution functions~\cite{Efremov:2002ut,Gamberg:2003pz,Mao:2012dk,Lu:2014fva}.
However, quantitative effects of the twist-3 T-odd fragmentation on the longitudinal SSAs have never been considered, because of the poor knowledge of $\tilde{G}^\perp(z,k_T^2)$.
For this reason we will study the two twist-3 T-odd fragmentation functions $G^\perp$ and $\tilde G^\perp$, using a spectator model, which assumes that the final hadron $h$ is produced through the process $q\rightarrow h\,q^\prime$, with $q^\prime$ is the spectator quark.
The spectator model has also been applied to calculate the Collins functions of pion and kaon~\cite{Bacchetta:2001di,Bacchetta:2002tk,Gamberg:2003eg,Bacchetta:2003xn,Amrath:2005gv, Bacchetta:2007wc}, as well as the the twist-3 fragmentation functions $\tilde{H}(z,k_T^2)$ and $H(z,k_T^2)$.

\section{Model calculation of $G^{\perp}$ and $\tilde{G}^{\perp}$}


For the TMD fragmentation functions, one can define the following quark-quark correlation function~\cite{Bacchetta:2006tn}:
  \begin{multline}
\Delta(z,k_T)  =\frac{1}{2z}\sum_X \, \int
  \frac{d\xi^+  d^2\bm{\xi}_T}{(2\pi)^{3}}\;e^{i k \cdot \xi}\,
    \langle 0|\, {\cal U}^{\infty^+}_{(\bm{\infty}_T,\bm\xi_T)} {\cal U}^{\bm\xi_T}_{({\infty}^+,\xi^+)}
\,\psi(\xi)|h, X\rangle
\langle h, X|
             \bar{\psi}(0)\,
{\cal U}^{\bm{0}_T}_{(0^+,{\infty}^+)} {\cal U}^{\infty^+}_{(\bm{0}_T,\bm{\infty}_T)}
|0\rangle \bigg|_{\xi^-=0}\,.
\label{eq:delta}
\end{multline}
where the light-cone coordinates $a^{\pm} = a\cdot n_{\pm} = (a^0\pm a^3)/\sqrt{2}$ have been applied, $k$ and $P_h$ denote the momenta of the parent quark and produced hadron, respectively, and $k^{-}={P_h^{-}/{z}}$.
The notation ${\cal U}^{c}_{(a,b)}$ represents the Wilson line (gauge link) running along the direction from $a$ to $b$ at the fixed position $c$.
The detailed discussion on the Wilson line $\cal U$ has been given in Ref.~\cite{Bacchetta:2007wc}.
In the rest of this paper, we will utilize the Feynman gauge, in which the transverse gauge links ${\cal U}^{ \bm \xi_T}$ and ${\cal U}^{\bm 0_T}$  can be neglected~\cite{Ji:2002aa,Belitsky:2002sm}.

\begin{figure*}
  \includegraphics[width=0.4\columnwidth]{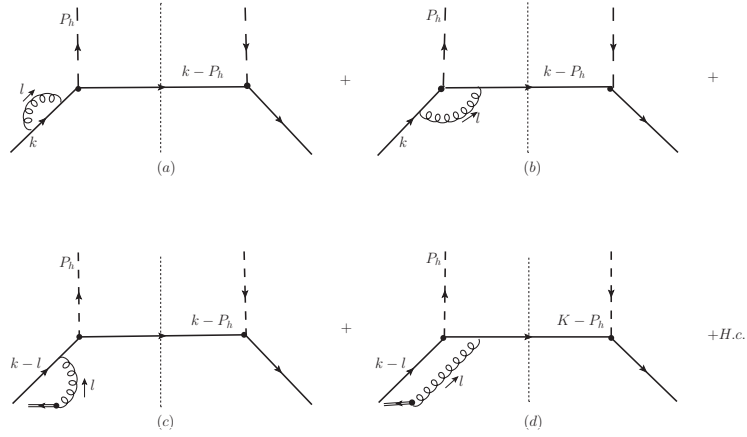}
 \caption {One loop order corrections to the fragmentation function of a quark into a meson in the spectator model. The double lines in (c) and (d) represent the eikonal lines.
Here ``H.c." stands for the hermitian conjugations of these diagrams.}
 \label{deltadiag}
\end{figure*}

Up to the twist-3 level, the fragmentation correlation function for a spinless hadron can be parameterized as
\begin{align}
 \label{eq:corr}
\Delta(z, k_T) &=
\frac{1}{2} \, \biggl\{ D_1 \nslash_- + \ii H_1^\perp \frac{ \bigl[
  \kslash_T, \nslash_- \bigr]}{2M_h}\biggr\}
\nonumber \\ &\quad
+ \frac{M_h}{2 P_h^-}\,\biggl\{ E +D^\perp \frac{\kslash_T}{M_h}+ \ii H
\frac{ \bigl[\nslash_-, \nslash_+ \bigr]}{2} + G^\perp
\g_5\,\frac{\eps_T^{\rho \sigma} \g_{\rho}\slim k_{T \sigma}}{M_h}\biggr\}.
\end{align}
As one can see, there are two T-odd fragmentation functions at twist-3.
One is $H(z, k_T^2)$, which has been studied in Ref.~\cite{Lu:2015wja} (its collinear version $H(z)$ has also been studied in Ref.~\cite{Kanazawa:2014tda}).
The other is the fragmentation function $G^\perp$.
Similar to the T-odd distribution function $g^\perp$,  $G^\perp$ appears in the parametrization in Eq.~(\ref{eq:corr}) because the direction of the Wilson line (e.g., $n_+$ in SIDIS) provides a vector independent of $P_h$ and $k$ for a Lorentz invariant decomposition of the correlator $\Delta(k)$~\cite{Bacchetta:2004zf,Goeke:2005hb}.
Specifically, $G^\perp$ is associated with the correlation $\gamma_5 \epsilon^{\mu\nu\rho\sigma} \gamma_\mu P_{h\nu} n_{+\rho} k_\sigma $.
From Eq.~(\ref{eq:corr}), $G^\perp(z, k_T^2)$ can be projected from the following trace
\begin{align}
{1\over P_h^-}\epsilon_T^{\alpha\beta} k_{T\beta} G^\perp(z,k_T^2) = {1\over 2}\textrm{Tr}[\Delta(z,k_T) \gamma^{\alpha}\gamma_5]\,. \label{eq:trace}
\end{align}
As is well known, the tree level calculation yields vanishing contributions to $G^\perp$, since the T-odd fragmentation functions are contributed by the imaginary part of the scattering amplitude~\cite{Amrath:2005gv}.
Therefore one has to consider the diagrams at the loop level.
In this paper, we will take into account the contribution of gluon rescattering at the one loop order, which is also essential to ensure the color gauge invariance of fragmentation functions.
The calculation of the fragmentation function $G^\perp$ in the spectator model is analogous to the previous calculations of the Collins function~\cite{Amrath:2005gv,Bacchetta:2007wc} and the twist-3 fragmentation function $H$~\cite{Lu:2015wja}.
In principle, there are four different diagrams (and their hermitian conjugates) may give rise to $G^\perp$, as shown in Fig.~\ref{deltadiag}.
However, in our case we find that the contributions from the self-energy diagram (Fig.~\ref{deltadiag}a) and the quark-meson vertex diagram (Fig.~\ref{deltadiag}b) vanish.
while the last two diagrams, the hard scattering vertex (Fig.~\ref{deltadiag}c) diagram and box diagram (Fig.~\ref{deltadiag}d), generate nonzero contributions.
In details, when we performed the trace in Eq.~(\ref{eq:trace}) using the correlator from Fig.\ref{deltadiag}a or Fig.~\ref{deltadiag}b, we obtain only one term which is proportional to $\epsilon^{\alpha\mu\nu\rho}k_\mu l_\nu P_{h\rho}$, with $l$ the loop momentum.
After the integration over $l$ is carried out, this term cannot give a contribution because of the property of the antisymmetric tensor $\epsilon$ (see also Eq.~(\ref{eq:intab})).
This is different from the calculation of the twist-3 chiral-odd fragmentation function $H$ for which Fig.\ref{deltadiag}a and Fig.~\ref{deltadiag}b give nonzero results.
The reason is that the Dirac structure  ($\sigma^{\alpha \beta}\gamma_5$) associated with the fragmentation function $H$ is more complicated the one associated with $G^\perp$, so that there are more terms appear besides the one proportional to $\epsilon^{\alpha\mu\nu\rho}k_\mu l_\nu P_{h\rho}$ after the trace is performed.

Using the spectator model, for the case of quark fragmenting into a pseudoscalar meson, we can write the expressions of the correlator contributed by Fig.~\ref{deltadiag}c and \ref{deltadiag}d as :
\begin{align}
 \begin{split}
\Delta_{(c)}(z,k_T) &=\ii\frac{4C_F\alpha_s }{2(2 \pi)^2(1-z)P_h^-}\, \frac{(\kslash + m)}{(k^2 -
  m^2)^2}\,  g_{qh}\gamma_5
\,(\kslash -\Pslash_h +m_s) g_{q h}  \gamma_5(\kslash +m)\\
&\quad\int\frac{d^4 l}{(2 \pi)^4} \,
\frac{ \gamma^- \,
(\kslash -
  \lslash +m)\, }{((k-l)^2 - m^2 +\ii \varepsilon) (-l^- \pm i
  \varepsilon)(l^2  +\ii \varepsilon)}\,,
\end{split} \displaybreak[0] \label{eq:deltac}\\
 \begin{split}
\Delta_{(d)}(z,k_T) &= \ii\frac{4C_F\alpha_s }{2(2 \pi)^2(1-z)P_h^-}\, \frac{(\kslash + m)}{k^2 -
  m^2}\,  g_{qh}\gamma_5
\,(\kslash -\Pslash_h +m_s)\\
&\quad\int\frac{d^4 l}{(2 \pi)^4} \,
\frac{ \gamma^- (\kslash - \Pslash_h - \lslash +m_s)\,
g_{q h}  \gamma_5\,
(\kslash -
  \lslash +m)\, }{((k-P_h-l)^2 -
  m_s^2+\ii \varepsilon ) ((k-l)^2 - m^2 +\ii \varepsilon) (-l^- \pm i
  \varepsilon)(l^2  +\ii \varepsilon)}\,,
\end{split}\label{eq:deltad}
\end{align}
where $g_{qh}$ is the coupling of the quark-hadron vertex, $m$ and $m_s$ represent the masses of the initial quark and the spectator quark, respectively.
Here $m_s$ is not constrained to be equal to $m$, following the choice in Refs.~\cite{Bacchetta:2007wc,Lu:2015wja}.
In Eqs.~(\ref{eq:deltac}) and (\ref{eq:deltad}) we have applied the Feynman rule  $1/{(-l^{-}\pm i\varepsilon)}$ for the eikonal propagator, as well as that for the vertex between the eikonal line and the gluon.
Note that the sign of the factor $\ii\varepsilon$ in the eikonal propagator is different for SIDIS $(+)$ and $e^+e^-$ annihilation $(-)$.
However, this will not affect the calculation of the fragmentation function $G^\perp$, which is similar to the case of the Collins function~\cite{Metz:2002iz,Collins:2004nx} and the twist-3 fragmentation function $H$.

To obtain the imaginary part of the correlator, we utilize the Cutkosky cut rule to put the gluon and quark lines on the mass shell.
This corresponds to the following replacements on the propagators by using the Dirac delta functions
\begin{align}
{1\over l^2 + \ii\varepsilon} \rightarrow -2\pi i\delta(l^2),~~~~~~~ {1\over (k-l)^2 + \ii\varepsilon} \rightarrow -2\pi i\delta((k-l)^2) \,.\label{eq:cuts}
\end{align}
The other combinations (cutting through the eikonal line or the spectator line) yield zero contributions, as shown in Refs.~\cite{Lu:2015wja,Yuan:2008it}.
Moreover, using the Cutkosky cut rule may also avoid dealing with the issue of renormalization~\cite{Bacchetta:2001di,Bacchetta:2002tk}.

Using the cut rules in Eq.(~\ref{eq:cuts}), we perform the integration over the loop momentum $l$ and arrive at the following schematic form for $G^\perp(z,k_T^2)$:
\begin{align}
G^\perp(z,k_T^2) & = {2 C_F\alpha_s g_{qh}^2\over (2\pi)^4 (1-z) }\,{1\over (k^2-m^2)}\,\left({G^\perp}_{(c)}(z,k_T^2)+{G^\perp}_{(d)} (z,k_T^2)\right)\,, \label{eq:Gzkt}
\end{align}
where the two terms ${G^\perp}_{(c)}(z,k_T^2)$ and ${G^\perp}_{(d)}$ on the r.h.s. of (\ref{eq:Gzkt}) correspond to the contributions from Fig.~\ref{deltadiag}c and Fig.~\ref{deltadiag}d:
\begin{align}
{G^\perp}_{(c)} (z,k_T^2) &= {2zI_{3}k^- + 2zI_1}\,,
 \label{eq:thc}\\
{G^\perp}_{(d)} (z,k_T^2) &= {2zI_{1}+2(k^2+m^2)\mathcal{C}+2(k^2+m(m-2m_s))\mathcal{D}
+2(1-z)\mathcal{E}P_h^- }\,.
\end{align}
Here $k^2 = z k_T^2/(1-z) + m_s^2/(1-z) + m_h^2/z $, and $\mathcal{C}$, $\mathcal{D}$ and $\mathcal{E}$ denote the following functions
\begin{align}
\mathcal{C} &={I_{34}k^-\over 2k_T^2} +{1\over 2zk_T^2}\left(-zk^2 + \left(2-z\right) m_h^2 + zm_s^2 \right)I_2, \\
\mathcal{D} &={-I_{34}k^-\over 2zk_T^2} -{1\over 2zk_T^2}\left(\left(1-2z\right)k^2 + m_h^2 - m_s^2 \right)I_2, \\
\mathcal{E} &={\lambda(m_h,m_s)\over 4zP_h^-k_T^2}I_{2} -{1\over 4z^2k_T^2}\left(\left(1-2z\right)k^2 + m_h^2 - m_s^2 \right)I_{34} +{{k^2-m^2}\over 2}I_4,
\end{align}
which originate from the integration
\begin{align}
\int d^4l { l^\mu\, \delta(l^2)\, \delta((k-l)^2-m^2)\over ((k-P_h-l)^2-m_s^2)(-l\cdot n_++i\epsilon)}
=\mathcal{C}\, k^\mu + \mathcal{D}\, P_h^\mu + \mathcal{E}n_+^\mu.
\end{align}
The functions $I_{i}$ in the above equations are defined as~\cite{Amrath:2005gv}
\begin{align}
I_{1} &=\int d^4l \delta(l^2) \delta((k-l)^2-m^2) ={\pi\over 2k^2}\left(k^2-m^2\right)\,, \\
I_{2} &= \int d^4l { \delta(l^2) \delta((k-l)^2-m^2)\over (k-P_h-l)^2-m_s^2}
={\pi\over 2\sqrt{\lambda(m_h,m_s)} }  \ln\left(1-{2\sqrt{ \lambda(m_h,m_s)}\over k^2-m_h^2+m_s^2 + \sqrt{ \lambda(m_h,m_s)}}\right)\,, \\
I_{3} & = \int d^4l {\delta(l^2) \delta((k-l)^2-m^2) \over -l^- +i\varepsilon}\,,\\
I_{4} & = \int d^4l {\delta(l^2) \delta((k-l)^2-m^2) \over (-l^- +i\varepsilon)((k-P_h-l)^2-m_s^2)}\,,
\end{align}
with $\lambda(m_h,m_s)=(k^2-(m_h+m_s)^2)(k^2-(m_h-m_s)^2)$,
and $I_{34}$ is the linear combination of $I_3$ and $I_4$,
\begin{align}
I_{34} = k^-\left(I_3+(1-z)(k^2-m^2)I_4\right) = \pi\ln{\sqrt{k^2}(1-z)\over m_s}\,,
\end{align}
As one can see, the two terms $G^\perp_{(c)}(z,k_T^2)$ and $G^\perp_{(d)}(z,k_T^2)$ are separately divergent, due to  the appearance of $I_3$ and $I_4$ in their expressions.
However, their sum is eventually finite.
Therefore, $G^\perp(z,k_T^2)$ is free of light-cone divergences in the spectator model calculation.
This is different from the calculation of the distribution function $g^\perp$~\cite{Gamberg:2006ru}, for which a light-cone divergence arises.
We note that the reason of this distinction is that the kinematical configuration responsible for the nonzero T-odd fragmentation functions is different from that for the T-odd distribution functions.
Finally, we arrive at the result of $G^\perp(z,k_T^2)$:
\begin{align}
{G^\perp}(z,k_T^2) &=  {2C_F \alpha_s g_{qh}^2\over (2\pi)^4 (1-z) }\,{1\over (k^2-m^2)}\,\bigg{\{}2z{I_1}+2\left(k^2+m^2\right)\mathcal{C}+2\left(k^2+m(m-2m_s)\right)\mathcal{D}\nonumber\\
  &\left.+{(1-z)\over zk_T^2}\left(\lambda(m_h,m_s)I_2+\left((1-2z)k^2+m^2_h-m_s^2\right)I_{34}\right)+2zI_{34}k^{-}\right\}\,. \label{eq:gtrs}
\end{align}

\begin{figure*}
  \includegraphics[width=0.4\columnwidth]{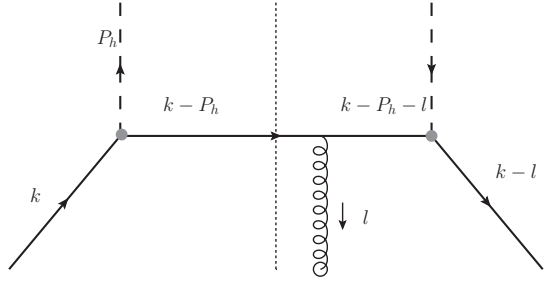}
 \caption {Diagram relevant to the calculation of the qgq correlator in the spectator model}
 \label{fig:qgqspec}
\end{figure*}

In the following, we present the calculation of the fragmentation function $\tilde{G}^\perp(z,k_T^2)$,
which can be projected from the following trace~\cite{Bacchetta:2006tn}:
\begin{align}
{z\over 2}\Tr[\tilde{\Delta}_{A\rho}(z,k_T) (g_T^{\alpha\rho}-i\epsilon_T^{\alpha\rho}\gamma_5)\gamma^-] = k_T^\alpha(\tilde{D}^\perp(z, k_T^2) - i \tilde{G}(z, k_T^2))\,. \label{eq:proj}
\end{align}
where $\tilde{\Delta}_{A}^{\alpha}(z,k_T) $ denotes the twist-3 quark-gluon-quark correlator for fragmentation, which is defined as~\cite{Bacchetta:2006tn,Metz:2012ct}:
\begin{align}
\tilde{\Delta}_A^\alpha(z,k_T) &=\sum_{X}\hspace{-0.55cm}\int \; \frac{1} {2z}\int \frac{d\xi^{+}d^2\bm\xi_T} {(2\pi)^3}\int  e^{\ii k\cdot \xi} \langle 0| \int^{\xi^+}_{\pm\infty^+} d{\eta^+}\mathcal{U}^{\bm\xi_T}_{(\infty^+,\eta^+)}\nonumber\\
 &\times gF^{-\alpha}_\perp (\eta) \mathcal{U}^{\bm\xi_T}_{(\eta^+,\xi^+)} \psi(\xi)|P_{h};X\rangle\langle P_{h};X|\bar{\psi}(0)\mathcal{U}^{\bm 0_T}_{(0^+,\infty^+)}\mathcal{U}^{\infty^+}_{(\bm 0_T,\bm \xi_T)}|0\rangle\bigg|_{\begin{subarray}{l}
\eta^+ = \xi^+=0 \\ \eta_T = \xi_T \end{subarray}}\,.
\label{eq:qgq}
\end{align}
Here $F^{\mu\nu}$ is the antisymmetric field strength tensor of the gluon.
The diagram used to calculate the fragmentation function $\tilde{G}^\perp$ in the spectator model is shown in Fig.~\ref{fig:qgqspec}.
The left and right hand sides of Fig.~\ref{fig:qgqspec} correspond to the quark-hadron vertex $\langle P_{h};X|\bar{\psi}(0)|0\rangle $ and the vertex containing gluon rescattering $\langle 0|igF_{\perp}^{-\alpha}(\eta^{+})\psi(\xi^{+})|P_{h};X\rangle$, respectively.
The expressions for these vertices in the spectator model can be easily obtained.
After considering all these ingredients, we can write down the expression of the quark-gluon-quark correlator
\begin{align}
 \begin{split}
\tilde{\Delta}_A^\alpha(z,k_T)  &= \ii\frac{C_F\alpha_s }{2(2 \pi)^2(1-z)P_h^-}\,{1\over k^2-m^2}\\
&\int\frac{d^4 l}{(2 \pi)^4} \,
\frac{(l^-g_T^{\alpha \mu} -l_T^\alpha g^{-\mu}) (\kslash-\lslash + m)\,  g_{qh}\gamma_5
\,(\kslash - \Pslash_h -\lslash +m_s) \gamma_\mu (\kslash - \Pslash_h +m_s)\,
g_{q h}  \gamma_5\,
(\kslash  +m)\, }{(-l^- \pm\ii
  \varepsilon)((k-l)^2 - m^2 -\ii \varepsilon) ((k-P_h-l)^2 -
  m_s^2-\ii \varepsilon ) (l^2  -\ii \varepsilon)}\,, \label{eq:qgqexp}
\end{split}
\end{align}
where the factor $(l^-g_T^{\alpha \mu} -l_T^\alpha g^{-\mu})$ comes from the Feynman rule corresponding to the gluon field strength tensor, as denoted by the open circle in Fig.~\ref{fig:qgqspec}.

Similar to ${G}^\perp$, the contribution to T-odd fragmentation function $\tilde{G}^\perp$ also originates from the imaginary part of the one-loop diagram~\cite{Bacchetta:2001di}.
Again, we can obtain the imaginary part by applying the Cutkosky cut rule and integrating over the internal momentum $l$, similar to the fragmentation function $G^\perp$. In fact, we consider every possible cuts on the propagators appearing in Fig.~\ref{fig:qgqspec}.
However, we find that only the cut on the gluon line and the fragmenting quark survive after the replacement in Eq.~(\ref{eq:cuts}), and the other choices are kinematically forbidden or cancel out each other~\cite{Lu:2015wja}.

Therefore, we apply again the cut rules given in Eq.~(\ref{eq:cuts}) to calculate the fragmentation function $\tilde{G}^\perp$.
After performing the integration over loop momentum $l$, the final result for $\tilde{G}^\perp$ is led to
\begin{align}
 \tilde{G}^\perp(z,k_T^2)& =
 -\frac{C_F\alpha_s g_{qh}^2}{(2 \pi)^4(1 - z)}\,\frac{1}{(k^2 - m^2)} \bigg{\{}\big((m-m_s)^2-m_h^2\big)\big{[}z\mathcal{A}+z\mathcal{B}-2I_{2}\nonumber\\
&+\left(2-\frac{2}{z}\right)\mathcal{C}\big{]}+4z(m^2\mathcal{A}+m(m-m_s)\mathcal{B})\nonumber\\
&+ (k^2-m^2){[}(z-1)\mathcal{C}-zI_2{]}-zI_1\bigg{\}}\,, \label{Gtilde}
\end{align}
where $\mathcal{A}$ and $\mathcal{B}$ denote the following functions
\begin{align}
\mathcal{A}&={I_{1}\over \lambda(m_h,m_s)} \left(2k^2 \left(k^2 - m_s^2 - m_h^2\right) {I_{2}\over \pi}+\left(k^2+m_h^2 - m_s^2\right)\right), \\
\mathcal{B}&=-{2k^2 \over \lambda(m_h,m_s) } I_{1}\left (1+{k^2+m_s^2-m_h^2 \over \pi} I_{2}\right),
\end{align}
which appear in the decomposition of the integral
 \begin{align}
\int d^4l { l^\mu\, \delta(l^2)\, \delta((k-l)^2-m^2)\over (k-P_h-l)^2-m_s^2}
=\mathcal{A}\, k^\mu + \mathcal{B}\, P_h^\mu. \label{eq:intab}
\end{align}

In principle, in calculating $\tilde{G}^\perp$, one could also consider a diagram where the upper vertex of the gluon in Fig.~\ref{fig:qgqspec} attaches to the quark leg on the r.h.s. (the one with momentum $k-l$).
In an explicit calculation we obtain the following result from this diagram
\begin{align}
k^\alpha \tilde{G}^{\perp}(z,k_T^2)& \propto \int {d^4 l \over (2\pi)^4} \,{k^\alpha l^- -k^- l^\alpha \over -l^-\pm i\varepsilon} \,\delta(l^2) \delta((k-l)^2-m^2). \label{eq:figcon}
\end{align}
Using the decomposition of the integral ($\mathcal{F}$ and $\mathcal{G}$ are the scalar functions of $k$, $n_+$ and $m$)
 \begin{align}
  \int d^4 l   \,{l^\mu \over -l\cdot n_+\pm i\varepsilon} \,\delta(l^2) \delta((k-l)^2-m^2)
  = \mathcal{F}k^\mu + \mathcal{G} n_+^\mu,
\end{align}
one can conclude that Eq.~(\ref{eq:figcon}) should give zero contribution to $\tilde{G}^{\perp}$.

\section{numerical results}

In this section, we present the numerical results of the fragmentation functions $G^{\perp,\pi^+/u}$ and $\tilde{G}^{\perp,\pi^+/u}$, which belong to the so-called favored fragmentation functions.
To do this we define the half-$k_T$ moment for them
\begin{align}
G^{\perp\,(1/2)}(z) &= \int d^2 \bm K_T {|\bm k_T|\over 2m_h}\,G^\perp(z,k_T^2) =  z^2\int d^2\bm k_T {|\bm k_T|\over 2m_h}\,G^\perp(z,k_T^2)\,, \\
\tilde{G}^{\perp\,(1/2)}(z) & = \int d^2\bm K_T{|\bm k_T|\over 2m_h}\,\tilde{G}^\perp(z,k_T^2)=
 z^2\int d^2\bm k_T {|\bm k_T|\over 2m_h}\,\tilde{G}^\perp(z,k_T^2)\,,
\end{align}
where $\bm K_T = -z \bm k_T$ is the transverse momentum of the produced hadron with respect to that of the parent.
In the case in which the quark-hadron coupling $g_{qh}$ is a constant, corresponding to a point-like coupling, these integrations are divergent.
There are two approaches to cut off the divergence.
One is to set an upper limit for $k_T$~\cite{Amrath:2005gv}, the other is to choose a Gaussian form factor depending on $k^2$ for $g_{q h}$~\cite{Gamberg:2003eg,Bacchetta:2007wc}, instead of a point-like coupling:
\begin{align}
g_{qh} \rightarrow g_{qh} {e^{-{k^2\over\Lambda^2}}\over z}\,.\label{eq:gaussian}
\end{align}
where $\Lambda^2= \lambda^{2}z^\alpha(1-z)^\beta$, with $\lambda$, $\alpha$, and $\beta$ are the parameters of the model.
In this work we will adopt the second approach, that is, to use Eq.~(\ref{eq:gaussian}) as the quark-pion coupling in order to effectively cut off the high $k_T$ region.
This choice is based on phenomenological motivation, as it can reasonably reproduce the unpolarized fragmentation function~\cite{Bacchetta:2007wc}.
Note that if one uses Eq.(\ref{eq:gaussian}), one of the form factors should depend on the loop momentum $l$.
However, since the form factor is introduced to cut off the divergence, here we make a reasonable choice that the form factor only depends on $k^2$ to simplify the calculation.
The same choice has also been applied in Ref.~\cite{Bacchetta:2007wc,Lu:2015wja}.
The values for the parameters are taken from Refs.~\cite{Bacchetta:2007wc,Lu:2015wja}:
\begin{align}
&  \lambda=2.18~\textrm{GeV},  ~\alpha=0.5\,(\textrm{fixed}),  ~\beta=0\,(\textrm{fixed}),\nonumber\\
& g_{qh}=g_{q\pi}=5.09, ~ m=0.3~\textrm{GeV}\,(\textrm{fixed}), ~m_s=0.53~\textrm{GeV} \,, \nonumber
\end{align}
which are obtained from simultaneously fitting the model results of the unpolarized fragmentation function $D_1$ and the Collins function with the known parametrization~\cite{deFlorian:2007aj,Anselmino:2013vqa}.
For the pion mass we adopt $m_h = m_\pi = 0.135\,\textrm{GeV}$.
Finally, we choose the value of the strong coupling constant $\alpha_s=0.2$, following the choice in Ref.~\cite{Bacchetta:2007wc}.
As shown in Ref.~\cite{Bacchetta:2007wc}, this choice is phenomenologically successful in reproducing the azimuthal asymmetry in $e^+ e^-$ annihilation contributed by the Collins function\cite{Abe:2005zx}.

\begin{figure}
  \includegraphics[width=0.5\columnwidth]{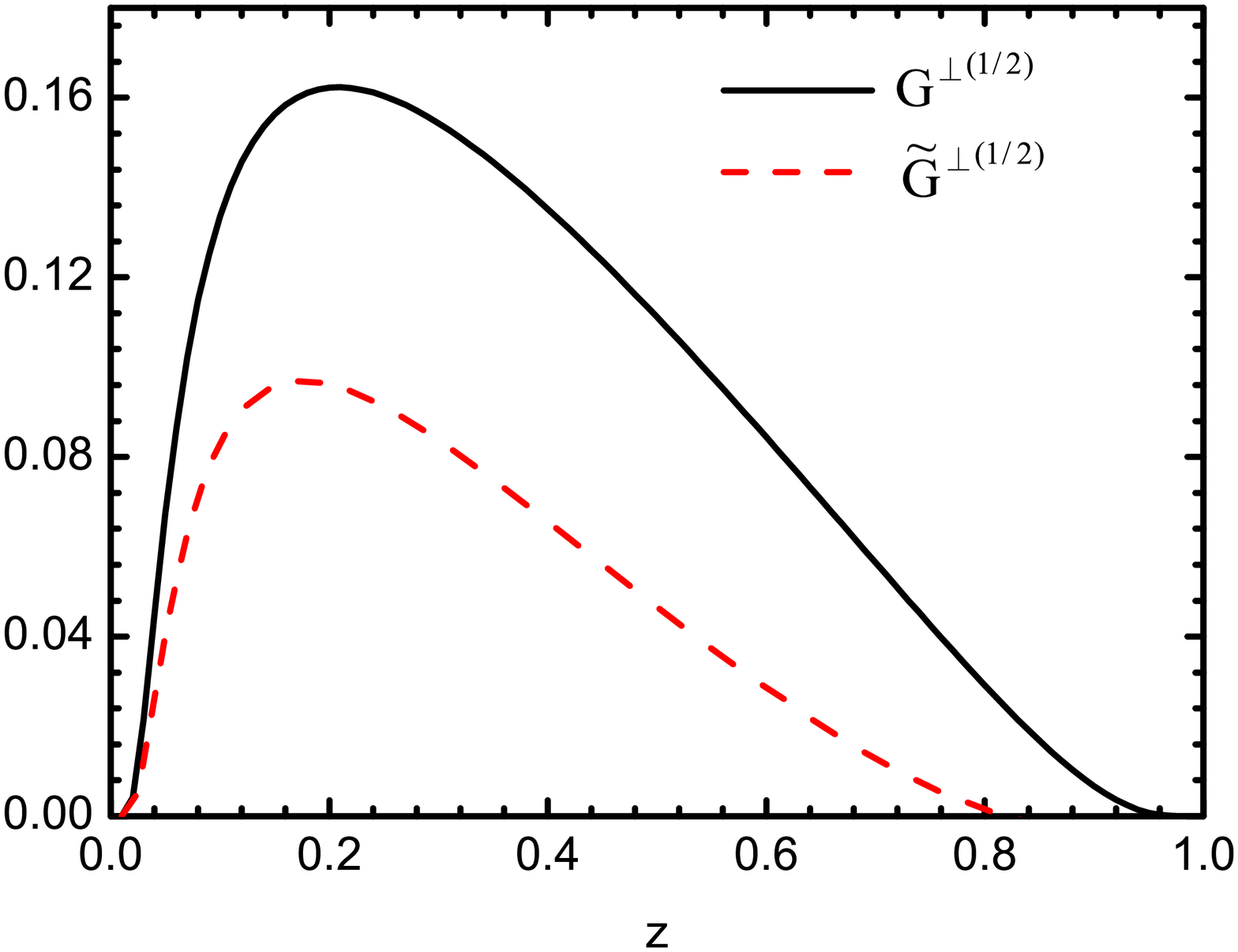}~~
  \includegraphics[width=0.5\columnwidth]{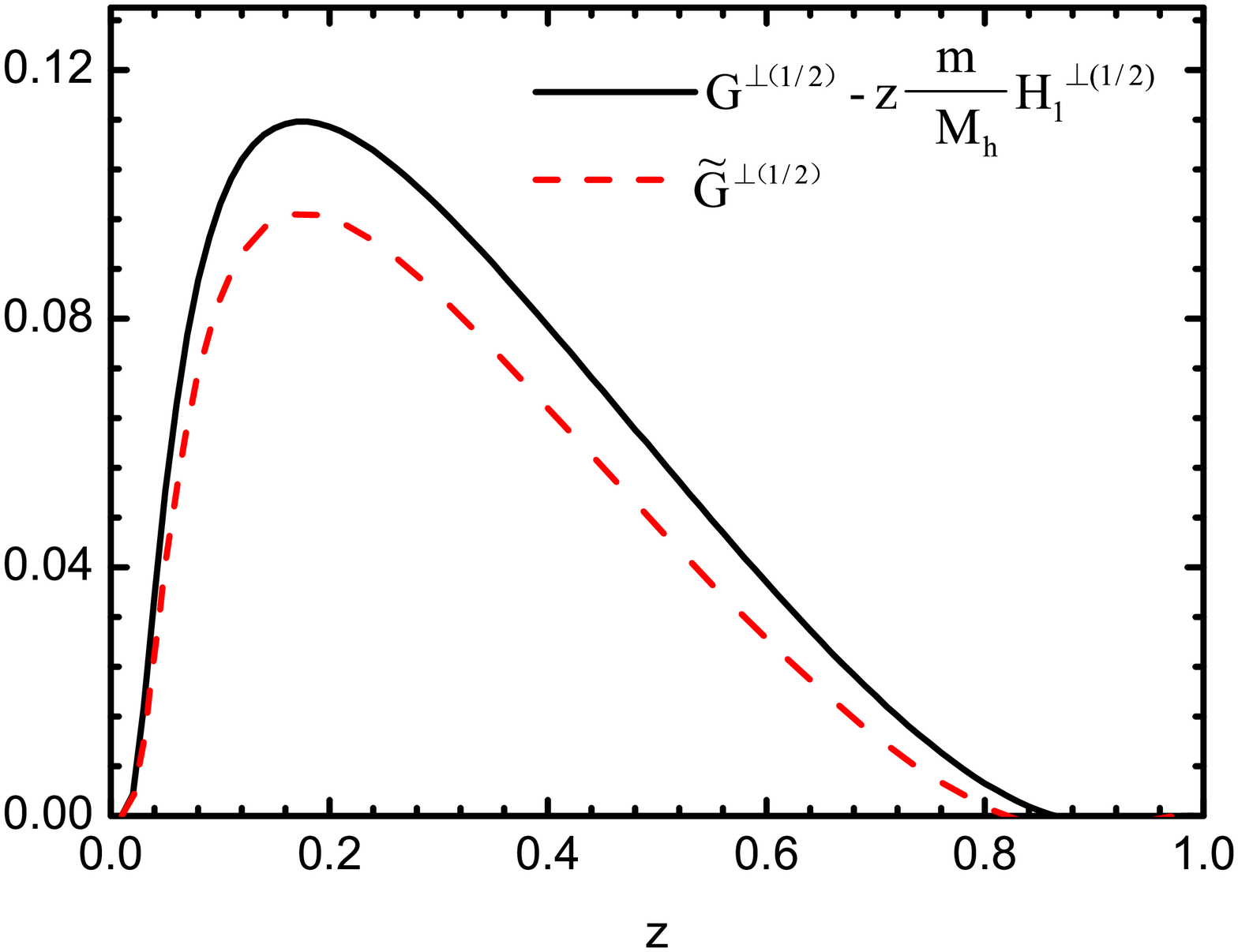}
 \caption {Left panel: The twist-3 fragmentation functions $G^{\perp(1/2)}(z)$ and $\tilde{G}^{\perp\,(1/2)}(z)$ vs $z$ in the spectator model, respectively.
Right panel: $\tilde{G}^{\perp\,(1/2)}(z)$ compared with $G^{\perp\,(1/2)}(z) - z{m\over M_h}H^{\perp\,(1/2)}_1(z)$ in the spectator model.}
\label{fig:Gtz}
\end{figure}

The numerical results for $G^{\perp\,(1/2)}(z)$ and $\tilde{G}^{\perp\,(1/2)}(z)$ as functions of $z$ are plotted in the left panel of Fig.~\ref{fig:Gtz}.
The dashed and solid lines represent $G^{\perp\,(1/2)}(z)$ and $\tilde{G}^{\perp\,(1/2)}(z)$ for the case $u~\rightarrow \pi^+$, respectively.
We find that their magnitudes are both sizable, although the size of $\tilde{G}^{\perp\,(1/2)}(z)$ is smaller than that of $G^{\perp\,(1/2)}(z)$.
Besides, they have a similar $z$-dependence, e.g., they are both positive and peak at around $z=0.2$
Considering that we chose a rather small value of $\alpha_s$, we point out that the magnitudes of $G^{\perp\,(1/2)}(z)$ and $\tilde{G}^{\perp\,(1/2)}(z)$ could be larger if a larger $\alpha_s$ is chosen.
Since $\tilde{G}^\perp$ is a chiral-even function, in principle it can couple to the unpolarized distribution $f_1$ in SIDIS to generate the $A_{UL}^{\sin\phi}$ and $A_{LU}^{\sin\phi}$ SSAs in SIDIS~\cite{Bacchetta:2006tn}.
Our numerical results on $\tilde{G}^\perp$ imply that twist-3 fragmentation contributions to these SSAs are non-negligible.

Using the QCD equation of motion for the quark fields, as well as the operator definitions of TMD fragmentation functions, one can derive the following relation among ${G}^\perp$, $\tilde{G}^\perp$ and the Collins function~\cite{Bacchetta:2006tn}:
\begin{align}
{G^{\perp}\over z}={\tilde{G}^{\perp}\over z}+{m\over M_h}H^{\perp}_1. \label{eq:eom}
\end{align}
On the one hand, the above relation demonstrates that the three fragmentation functions are not independent.
On the other hand, it may be used to verify the validity of the model calculation.
Due to the model calculations for them come from different Feynman diagrams, we cannot obviously find the relation from Eqs.(\ref{eq:Gzkt}), (\ref{Gtilde}) and the expression of $H_1^\perp$.
Therefore, we check numerically if the equation of motion relation (\ref{eq:eom})) holds in our model calculation.
In the right panel of Fig.~\ref{fig:Gtz}, we plot $\tilde{G}^{\perp\,(1/2)}(z)$ (dashed line) and $G^{\perp (1/2)}(z)-z{m\over M_h}H_1^{\perp\,(1/2)} (z)$ (solid line) vs $z$,
with $H_1^{\perp\,(1/2)} (z)$ the half $k_T$-moment of the Collins function:
\begin{align}
H_1^{\perp\,(1/2)}(z)= z^2\int d^2\bm k_T {|\bm k_T|\over 2M_h}\,H_1^\perp(z,k_T^2)\,,
\end{align}
By comparing $\tilde{G}^{\perp\,(1/2)}(z)$ and $G^{\perp\,(1/2)}(z)-z{m\over M_h}H_1^{\perp\,(1/2)} (z)$ in Fig.~\ref{fig:Gtz}, we find that the relation holds approximately in the model.
Thus, this result also provides a crosscheck on the validity of our calculation.

\section{Conclusion}

In this paper, we have studied the twist-3 T-odd fragmentation function $G^\perp (z, k^2_T)$ and $\tilde G^\perp(z, k_T^2)$ by considering the gluon rescattering effects.
We have employed the spectator model to compute all the possible diagrams contributing to
$G^\perp$.
We find that the two diagrams containing eikonal propagators give rise to nonzero $G^\perp$, and
the final result is free of light-cone divergence.
The function $\tilde G^\perp(z, k_T^2)$ is also calculated by exploiting the quark-gluon-quark correlator in the spectator model.
Using a Gaussian form factor for the quark-hadron vertex, we have estimated the half-$k_T$ moments of $G^\perp$ and $\tilde{G}^\perp$.
we find that their sizes are substantial in the spectator model, which implies that $\tilde{G}^\perp$ may provide considerable contributions to the longitudinal beam or target SSAs at the twist-3 level.
We have also verified numerically the equation of motion relation among $G^\perp$, $\tilde{G}^\perp$ and $H_1^\perp$, and find that it holds approximately in our calculation.
Our study may provide useful information on the twist-3 T-odd fragmentation functions, as well as their role in the SSAs in SIDIS.

\section*{Acknowledgements}
This work is partially supported by the National Natural Science
Foundation of China (Grants No.~11575043 and No.~11120101004), by the Qing Lan Project, and by Fondecyt (Chile) grants 1140390 and FB-0821.

\end{document}